\documentclass{article}
\usepackage[a4paper, total={6in, 10in}]{geometry}
\usepackage[svgnames,table]{xcolor}
\usepackage[utf8]{inputenc}
\usepackage[vlined,ruled]{algorithm2e}
\usepackage{amsmath}
\usepackage{amssymb}
\usepackage{bm,array}
\usepackage{booktabs}
\usepackage{enumitem}
\usepackage{flafter}
\usepackage{graphicx}
\usepackage{hyperref}
\usepackage{makecell}
\usepackage{multirow}
\usepackage{tabularx}
\usepackage{nicefrac}
\usepackage{url}
\usepackage{xspace}
\usepackage{perpage} 
\MakePerPage{footnote} 

\hypersetup{
	pdftitle={Unary and Binary Classification Approaches and their Implications for Authorship Verification},
	pdfauthor={Oren Halvani, Christian Winter, Lukas Graner}, 
	pdfsubject={Authorship Verification}, 
	pdfcreator={}, 
	pdfproducer={}, 
	pdfkeywords={Authorship Verification},
	pdfpagemode=UseNone,
	pdfstartview={XYZ null null null},
	colorlinks = true,
	linkcolor = red,
	citecolor = blue,
	urlcolor = blue,
	anchorcolor = blue
}

\definecolor{Unary}{RGB}{217,240,211}
\definecolor{BinaryIntrinisc}{RGB}{231,212,232}
\definecolor{BinaryExtrinisc}{RGB}{254,224,144}
\newenvironment{myitemize}   {\begin{itemize}   \setlength{\itemsep}{0pt}} {\end{itemize}}
\newenvironment{myenumerate} {\begin{enumerate} \setlength{\itemsep}{0pt}} {\end{enumerate}}

\newcommand*{\A}            {\mathcal{A}}
\newcommand*{\notA}         {\neg\A}
\newcommand*{\unknown}      {\mathcal{U}}

\newcommand*{\D}            {\mathcal{D}}
\newcommand*{\DA}           {\D_{\A}}

\newcommand*{\Dunk}         {\D_{\,\unknown}}

\newcommand*{\Dset}         {\mathbb{D}}

\newcommand*{\Corpus}       {\mathcal{C}}

\newcommand*{\CorpusAmazon}      {\Corpus_{\textrm{Amazon}}}
\newcommand*{\CorpusReddit}      {\Corpus_{\textrm{Reddit}}}
\newcommand*{\CorpusAmazonTrain} {\Corpus_{\textrm{Amazon}}^{\textrm{(tr)}}}
\newcommand*{\CorpusRedditTrain} {\Corpus_{\textrm{Reddit}}^{\textrm{(tr)}}}

\newcommand*{\Model}        {\mathcal{M}}

\newcommand*{\Arefset}        {\Dset_{\A}}
\newcommand*{\Problem}        {\rho}

\newcommand*{\Threshold}      {\theta}
\newcommand*{\ThresholdModel} {\theta_{\Model}}

\newcommand*{\occav}         {$\mathsf{OCCAV}$\mbox{}\xspace}
\newcommand*{\mocc}          {$\mathsf{MOCC}$\mbox{}\xspace}
\newcommand*{\aveer}         {$\mathsf{AVeer}$\mbox{}\xspace}
\newcommand*{\coav}          {$\mathsf{COAV}$\mbox{}\xspace}
\newcommand*{\veenmanNNCD}   {$\mathsf{NNCD}$\mbox{}\xspace}
\newcommand*{\glad}          {$\mathsf{GLAD}$\mbox{}\xspace}
\newcommand*{\cng}           {$\mathsf{CNG}$\mbox{}\xspace}
\newcommand*{\koppelGI}      {$\mathsf{GenIM}$\mbox{}\xspace}
\newcommand*{\khonjiGI}      {$\mathsf{ASGALF}$\mbox{}\xspace}
\newcommand*{\koppelUnmask}  {$\mathsf{Unmasking}$\mbox{}\xspace}

\newcommand*{\stamatatosProf}{$\mathsf{ProfileAV}$\mbox{}\xspace}
\newcommand*{\noeckerDist}   {$\mathsf{DistAV}$\mbox{}\xspace}
\newcommand*{\occKNN}        {$\mathsf{OCNN}$\mbox{}\xspace}

\newcommand*{\occSVM}        {$\mathsf{OSVM}$\mbox{}\xspace}
\newcommand*{\occLOF}        {$\mathsf{LOF}$\mbox{}\xspace}
\newcommand*{\occIF}         {$\mathsf{IF}$\mbox{}\xspace}

\renewcommand*{\AA}     {AA\mbox{}\xspace}

\renewcommand*{\aa}     {authorship attribution\mbox{}\xspace}

\newcommand*{\AV}       {AV\mbox{}\xspace}

\newcommand*{\av}       {authorship verification\mbox{}\xspace}

\newcommand*{\classY}      {\texttt{Y}\mbox{}\xspace}
\newcommand*{\classN}      {\texttt{N}\mbox{}\xspace}

\newcommand*{\accuracy}    {Accuracy\mbox{}\xspace}
\newcommand*{\fOne}        {F$_{1}$\mbox{}\xspace}

\newcommand*{\auc}         {AUC\mbox{}\xspace}

\newcommand*{\eg}            {e.\,g.,\mbox{}\xspace}
\newcommand*{\ie}            {i.\,e.,\mbox{}\xspace}
\newcommand*{\vs}            {versus\mbox{}\xspace}

\newcommand{\e}[1]{\emph{#1}}    

\begin{document}
\title{Unary and Binary Classification Approaches \\and their Implications for Authorship Verification}
\author{Oren Halvani, Christian Winter, Lukas Graner}

\date{\vspace{-5ex}}
\maketitle

\begin{abstract}
Retrieving indexed documents, not by their topical content but their writing style opens the door for a number of applications in information retrieval (IR). One application is to retrieve textual content of a certain author X, where the queried IR system is provided beforehand with a set of reference texts of X. Authorship verification (AV), which is a research subject in the field of digital text forensics, is suitable for this purpose. The task of AV is to determine if two documents (i.e. an indexed and a reference document) have been written by the same author X. Even though AV represents a unary classification problem, a number of existing approaches consider it as a binary classification task. However, the underlying classification model of an AV method has a number of serious implications regarding its prerequisites, evaluability, and applicability. In our comprehensive literature review, we observed several misunderstandings regarding the differentiation of unary and binary AV approaches that require consideration. The objective of this paper is, therefore, to clarify these by proposing clear criteria and new properties that aim to improve the characterization of existing and future AV approaches. Given both, we investigate the applicability of eleven existing unary and binary AV methods as well as four generic unary classification algorithms on two self-compiled corpora. Furthermore, we highlight an important issue concerning the evaluation of AV methods based on fixed decision criterions, which has not been paid attention in previous AV studies.
\end{abstract}

\section{Introduction} \label{Introduction}
Document classification plays a vital role across numerous fields of studies including library, information or computer science and represents a major task in IR. The categorization of documents\footnote{Note that ``documents'' are not necessarily restricted to natural language, but might also represent source code snippets or other types of textual data.} can be performed regarding a variety of concepts such as \textbf{genre}, \textbf{register}, \textbf{text type}, \textbf{domain}, \textbf{sublanguage} or \textbf{style}. Focusing on the (writing) style of documents allows a number of promising applications in IR. For example, a user can provide an IR system reference texts of an author $\A$ such as blog posts and ask the system to retrieve additional content of $\A$ (\eg product reviews, articles or comments) based on the same writing style. Style-based IR is in particular interesting if both reference and indexed documents stem from the same person but differ in terms of meta data (for instance, different user names) or are not even provided with meta data at all. This might be a helpful supplement in the context of fake news detection.

A number of research disciplines exist that concern themselves with the analysis of writing style (more precisely, with the authorship of documents), where the most important are \textbf{\aa} (\AA) and \textbf{\av} (\AV). The former deals with the problem to identify the most likely author of an unknown document $\Dunk$, given a set of texts of candidate authors. \AV, on the other hand, focuses on the question\footnote{There are also other formulations that describe \av problems (see, for example, \cite{SteinMetaAnalysisAV:2008}).} if $\Dunk$ was in fact written by a known author $\A$, where only a set $\Arefset$ of reference texts of this  author is given. Both disciplines are strongly related to each other, as any \AA problem can be broken down into a series of \AV problems \cite{StamatatosPothaImprovedIM:2017}. Here, an \AV system must determine for each verification problem $\rho = (\Dunk, \Arefset)$, if all involved documents stem from the same author, based on a specific \textbf{decision criterion}. Breaking down an \AA problem into multiple \AV problems is especially important in such scenarios, where the presence of $\Dunk$'s true author in the candidate set cannot be guaranteed. In contrast to \AA, which represents an $\bm{n}$\textbf{-ary} (multi-class) classification problem, \AV is a \textbf{unary} (one-class) classification\footnote{In this form, as highlighted in \cite{GLAD:2015}, \AV is considered to be a \textbf{recognition} rather than a \textbf{classification} problem.} problem, as there is only one class ($\A$) to learn from \cite{GLAD:2015,KoppelAVOneClassClassification:2004,StamatatosProfileCNG:2014,PothaStamatatosTopicAV:2018,SteinMetaAnalysisAV:2008}. However, inspecting previous studies reveals that unary classification appears to be a gray area in machine learning and, in particular, in the context of \AV, where a number of misunderstandings can be observed.

The \textbf{objective} of this paper is to analyze these misunderstandings in detail and to propose \textbf{clear criteria and properties} that aim to \textbf{close the gap of existing definitions} and attempts to characterize \AV methods, especially regarding the underlying classification models. By this, we hope to contribute to the further development of this young\footnote{According to the literature \cite{PANOverviewAV:2014}, Stamatatos et al. were the first researchers, who discussed \AV in the context of natural language texts in their paper \cite{StamatatosAutomaticTextCategorization:2000} published in 2000. \AV, therefore, can be seen as a young field in contrast to \AA, which dates back to the 19th century \cite{HolmesEvolutionStylometryHumanities:1998}.} research field. Based on our definitions, we investigate the applicability of eleven existing unary and binary \AV methods as well as four generic unary classification algorithms on two self-compiled corpora, which we make available for the \AV community. Furthermore, we elaborate the \textbf{implications} that have to be faced for each approach and highlight an important issue concerning the \textbf{evaluation} of such \AV methods that are based on fixed decision criterions.

\section{Existing Approaches} \label{RelatedWork}
In order to design an \AV method, a wide spectrum of possibilities exists including unary, binary and $n$-ary classification approaches. In the following subsection, we first describe a number of generic unary classification algorithms that \textbf{can} and \textbf{have} been used in the context of \AV. Afterwards, we present existing \AV approaches that were partially motivated by these algorithms. All introduced approaches will be assessed regarding their performance in Sec.~\ref{Evaluation}, where each approach will be categorized according to our proposed criteria and properties. 


\subsection{Generic Unary Classification Algorithms}
As can be observed in the literature (for example, \cite{HalvaniOCCAV:2018,HalvaniARES:2014,JankowskaAVviaCNG:2014,NealAVviaIsolationForests:2018}), a number of existing \AV methods are based on \textbf{unary classification algorithms}. In the following, we therefore provide a brief overview of some selected approaches, which were also considered in our evaluation.

\subsubsection{One-Class Nearest Neighbor}
One very simple, but quite effective, unary classification algorithm is \occKNN (\e{One-class Nearest Neighbor} \cite{TaxOCC:2001}). Given a unknown document $\Dunk$, the known documents $\Arefset = \{ \D_1, \D_2, \ldots ,\D_n \} $ and a predefined distance function, the idea behind \occKNN is that $\Dunk$ is accepted as a member of the target class $\A$, if its closest neighbor $\D_i$ within $\Arefset$ is closer to $\Dunk$ than the closest neighbor of $\D_i$ within $\Arefset \setminus \{\D_i\}$. A number of existing \AV approaches including \cite{HalvaniOCCAV:2018,HalvaniARES:2014,NoeckerDistractorlessAV:2012} represent modifications of \occKNN.

\subsubsection{One-Class Support Vector Machine (OSVM)}
The idea of \occSVM (\e{One-class Support Vector Machine} \cite{KhanOCCTaxonomy:2014}) is to construct a hypersphere shaped decision boundary with minimal volume around samples of $\A$ in a specific feature space and, by this, to distinguish all other possible documents of an unknown authorship. If $\Dunk$ falls inside this hypersphere, $\Dunk$ is accepted ($\unknown = \A$), otherwise rejected. In existing \AV and \AA works, \occSVM served as a baseline (for example, \cite{BoukhaledProbabilisticAV:2014,KoppelAVOneClassClassification:2004}) or as the core method \cite{NovinoSingleClassAA:2015}.

\subsubsection{Local Outlier Factor}
Another unary classification algorithm, which originally was constructed for finding outliers in large databases is \occLOF (\e{Local Outlier Factor} \cite{LOFBreunigKriegel:2000}) which, similarly to \occKNN, also employs nearest neighbor distances. \occLOF uses a sophisticated strategy for comparing the distances between $\Dunk$ and its $k$-nearest neighbors to the distances between these neighbors and their $k$-nearest neighbors. The final score \occLOF returns is a quotient, derived from all these distances, which becomes larger the more $\Dunk$ is an outlier.

\subsubsection{Isolation Forest}
Another unary classifier, which gained much attention in recent years, is \occIF (\e{Isolation Forest} \cite{IsolationForestLiu:2008}). Similarly to its counterpart \e{Random Forest}\footnote{Random Forest is a well-known classification algorithm, widely used for $n$-ary classification tasks.}, \occIF builds multiple binary trees that separate the feature space recursively. Each node divides its child nodes based on a randomly selected feature and threshold. Assuming that outliers are only ``few and different'' \cite{IsolationForestLiu:2008}, the idea is that instances placed deeper in a tree are less likely outliers. The acceptance of $\Dunk$ to be a member of the target class $\A$ depends on its corresponding depth, averaged over the trees. Neal et al. \cite{NealAVviaIsolationForests:2018} proposed an \AV approach that works on top of \occIF.

\subsection{Existing AV Approaches} \label{ExistingAVApproaches}
During 2013--2015, the organizers of the PAN\footnote{PAN is a series of scientific events and shared tasks on digital text forensics.} workshop held three \AV competitions \cite{PANOverviewAV:2013,PANOverviewAV:2015,PANOverviewAV:2014}, which attracted attention among the \AV community and led to a noticeable increase of proposed approaches in this field. In the following, we give a brief overview regarding a number of existing \AV methods which, at least partially, achieved promising results within the PAN competitions.

In 2013, Seidman \cite{SeidmanPAN13:2013} proposed a successful \AV method named \koppelGI (\e{General Impostors Method}) which is a slight variation of the well-known \e{Impostors} approach introduced by Koppel and Schler \cite{KoppelWinter2DocsBy1:2014}. \koppelGI works in two steps. First, so-called \textbf{impostor documents} are gathered that aim to represent the counter class of $\A$, namely $\notA$. Second, a \textbf{feature randomization} technique is applied iteratively to measure the similarity between pairs of documents. If, given this measure, a suspect is picked out from among the impostor set with sufficient salience, then the questioned document $\Dunk$ is considered to be written by this author, otherwise not \cite{KoppelWinter2DocsBy1:2014}. \koppelGI was the overall winning approach of the PAN-2013 \AV competition \cite{PANOverviewAV:2013} in terms of \fOne and was ranked second in terms of \auc. In 2014, Khonji and Iraqi \cite{KhonjiIraqiAV:2014} proposed a slightly modified version of \koppelGI, which they named \khonjiGI\footnote{\khonjiGI stands for ``A Slightly-modified GI-based Author-verifier with Lots of Features'' \cite{KhonjiIraqiAV:2014}.}. The authors adapted a modified min-max$(\cdot)$\footnote{Also known as the \e{Ruzicka} measure.} similarity measure as well as a larger set of features including function words, word shapes, and part-of-speech tags. \khonjiGI was the overall winning approach of the PAN-2014 \AV competition \cite{PANOverviewAV:2014}.

In 2015, Bagnall \cite{BagnallRNN:2015} proposed a method based on a character-level RNN, which was not only the overall winning approach at the PAN-2015 \AV competition \cite{PANOverviewAV:2015} but also the first attempt to apply deep learning in the context of \AV. Similarly to \koppelGI, the approach of Bagnall also \textbf{requires a corpus} $\Corpus = \{ \rho_1, \rho_2, \ldots, \rho_n \}$ with $\rho_i = ({\Dunk}_i, {\Arefset}_i)$, where the ratio of matching (\classY) and non-matching (\classN) authorships must be known beforehand. The method can be roughly split up into \textbf{four steps}. The \textbf{first step} is to train language models\footnote{Technically, language models considered in Bagnall's method are character probability distributions.} (LM) for all known document sets ${\Arefset}_1, {\Arefset}_2, \ldots, {\Arefset}_n$, which results in LM$_1$, LM$_2$, $\ldots$, LM$_n$. As a \textbf{second step}, each unknown document ${\Dunk}_i$ is attributed against all $n$ trained language models. For this, a score (more precisely, the \e{mean cross entropy}) between ${\Dunk}_i$ and each LM$_j$ is calculated, which describes how ``well'' LM$_j$ predicts ${\Dunk}_i$. In the \textbf{third step}, all scores across the problems are normalized, in order to overcome possible variances among the learned language models. In the \textbf{fourth step}, the normalized scores are ranked and transformed into similarity scores. Based on the \classY/\classN-ratio of $\Corpus$, the threshold to accept or reject unknown documents is then determined. For example, if $\Corpus$ is \textbf{balanced}\footnote{In a balanced corpus, the verification problems with matching (\classY) and non-matching (\classN) authorships are evenly distributed.}(which was the case for the PAN-2015 \AV corpora) then the method uses the median of the similarity scores as a threshold. One advantage of Bagnall's approach is that it considers the entire document as a sequence of characters and automatically learns patterns to distinguish between authors. By this, defining handcrafted features is avoided. However, since the method discriminates between multiple (known) authors, it better fits to the category of \AA instead of \AV methods (Dwyer \cite{DwyerAAviaLSTM:2017} also made this observation). This and the fact that we were not able to reproduce this complex approach led to our decision to exclude it from our evaluation. 

In 2015, Hürlimann et al. \cite{GLAD:2015} proposed their novel \AV approach \glad\footnote{\textbf{G}roningen \textbf{L}ightweight \textbf{A}uthorship \textbf{D}etection.}, which deliberately discards the idea to model an outlier class $\notA$ by collecting impostor documents. Instead, \glad considers a training corpus $\Corpus = \{ \rho_1, \rho_2, \ldots, \rho_n \}$, where each verification problem $\rho = (\Dunk, \Arefset)$ is labeled either as $\classY$ or $\classN$. Given $\Corpus$, \glad constructs for each $\rho in \in \Corpus$ (not each document in $\rho$) a feature vector consisting of 24 features. Here, the features were obtained individually from $\Dunk$, $\Arefset$ or simultaneously from both (Hürlimann et al. denote these as ``joint features''). After representing each $\rho_i$ in the feature space, the authors train a binary SVM to separate the space such that $\Dunk$ is accepted or rejected depending in which subspace it falls.

In 2018, Halvani et al. \cite{HalvaniOCCAV:2018} proposed their \AV method \occav\footnote{\textbf{O}ne-\textbf{C}lass \textbf{C}ompression \textbf{A}uthorship \textbf{V}erifier.}, which also avoids the idea to model a counter class $\notA$ and is even independent of a training corpus. \occav is inspired by the unary classification algorithm \occKNN. However, instead of constructing feature vectors from the documents, here, all texts are represented as compressed byte streams using the \e{Prediction by Partial Matching} (PPMd) algorithm. By this, a verification problem $\rho = (\Dunk, \Arefset)$ is transformed into $\hat\rho = (\expandafter\hat\Dunk, \expandafter\hat\Arefset)$, where all involved documents are compressed. Another difference to \occKNN is that instead of using a standard distance function, \occav relies on the so-called \e{Compression Based Cosine}, which measures dissimilarities between compressed documents. Given this measure and the compressed documents in $\hat\rho$, the method computes dissimilarities between $\expandafter\hat\Dunk$ and each $\expandafter\hat\DA \in \expandafter\hat\Arefset$. Next, $\hat\D_{\e{near}} \in \expandafter\hat\Arefset$ is selected, which has the smallest dissimilarity $d_{\e{min}}$ to $\expandafter\hat\Dunk$. Then, dissimilarities between $\hat\D_{\e{near}}$ and each $\hat\D_j \in \expandafter\hat\Arefset \setminus \{ \hat\D_{\e{near}} \}$ as well as their average $d_{\e{avg}}$ are computed. If $d_{\e{min}} < d_{\e{avg}}$ holds, then the unknown document $\Dunk$ is assumed to be written by $\A$.

\section{Analysis} \label{Analysis}
With the increasing number of proposed \AV approaches, the wish arose to compile a systematic characterization to enable a better comparison between the methods. In 2004, Koppel and Schler \cite{KoppelAVOneClassClassification:2004} described, for the first time, the connection between unary classification and \AV. In 2008, Stein et al. \cite{SteinMetaAnalysisAV:2008} provided an overview of important algorithmic building blocks for \AV where, among others, they also formulated three \AV problems as decision problems. In 2014, Potha and Stamatatos \cite{StamatatosProfileCNG:2014}, introduced specific properties that aim to characterize \AV methods. However, a deeper look on previous attempts to organize the field of \AV reveals a number of misunderstandings, in particular, when it comes to draw the borders between unary and binary \AV approaches. In the following, we analyze these misunderstandings and propose redefinitions as well as new \AV properties. We show that in fact there are not only two (unary/binary) but \textbf{three possible categories} of \AV methods and that their categorization solely depends on the way how the acceptance criterion is determined.

\subsection{Determinism of results} \label{AV_Property_Deterministic_Stochastic_Methods}
A fundamental property of any \AV method, especially in the context of evaluation, is whether it behaves deterministically or non-deterministically. \AV approaches such as \cite{GLAD:2015,JankowskaCNGAV:2013,StamatatosProfileCNG:2014} always generate the same output for the same inputs, \ie these methods are \textbf{deterministic}. In contrast, \textbf{non-deterministic} \AV methods as proposed in \cite{CastanedaAVviaLDA:2017,KoppelAVOneClassClassification:2004,NealAVviaIsolationForests:2018,StamatatosPothaImprovedIM:2017,SeidmanPAN13:2013} involve randomness (for instance, subsampling of the feature space or the number of impostors) which, as a consequence, might distort the evaluation since every run on a (training or test) corpus very likely leads to different results. Therefore, it is indispensable to perform multiple runs and to consider the average and dispersion of the achieved results for a reasonable and robust comparison between different \AV approaches.

\subsection{Optimizability}
Optimizability is another property of an \AV method, which affects the dependency on a training corpus. We define an \AV method as \textbf{optimizable} if, according to its \textbf{design}, it offers adjustable hyperparameters that can be tuned against a training corpus, given an optimization method (\eg grid or random search). Such hyperparameters might be, for instance, the selected distance/similarity function, the number of neurons/layers in a neural network, the chosen kernel method of an SVM, the selected feature categories, or adjustable weights and thresholds. The majority of existing \AV methods in the literature (including \cite{CastroAVAverageSimilarity:2015,EscalanteGomezParticleSwarmAV:2009,CastanedaAVviaLDA:2017,KoppelWinter2DocsBy1:2014,StamatatosProfileCNG:2014}) belong to this category. On the other hand, if a published \AV approach involves hyperparameters that have been entirely fixed such that there is no further possibility to improve its performance from outside (without deviating from the definitions in the publication of the method), the method is considered to be \textbf{non-optimizable}. Obviously, non-optimizable \AV approaches are easier to reproduce, as we can discard the dependency on a training corpus. Among the proposed \AV methods in the respective literature, we only identified three approaches \cite{HalvaniOCCAV:2018,NoeckerDistractorlessAV:2012,VeenmanPAN13:2013} that belong to this category.

\subsection{Model Category (Unary \vs Binary)}
Even though \AV clearly represents a unary classification problem \cite{GLAD:2015,KoppelAVOneClassClassification:2004,StamatatosProfileCNG:2014,PothaStamatatosTopicAV:2018,SteinMetaAnalysisAV:2008}, one can observe in the literature that sometimes it is interpreted as unary \cite{JankowskaAVviaCNG:2014,NoeckerDistractorlessAV:2012,NealAVviaIsolationForests:2018,StamatatosProfileCNG:2014} and sometimes as binary \cite{KocherSavoySpatiumL1:2017,KoppelWinter2DocsBy1:2014,LuyckxAVManyAuthorsLimitedData:2008,VeenmanPAN13:2013}. We define the way an \AV approach is modeled by the phrase \textbf{model category}. However, before explaining this in more detail, we first have to recall what, according to the literature, unary classification exactly represents. For this, we list the following verbatim quotes, which characterize unary classification, as can be seen, almost identically (emphasized by us):
\begin{myitemize}
	\item \e{``In one-class classification it is assumed that only information of one of the classes, the target class, is available. This means that \textbf{just example objects of the target class} can be used and that \textbf{no information about} the other class of \textbf{outlier objects is present}.''} \cite{TaxOCC:2001}

	\item \e{``One-class classification (OCC) [\dots] consists in \textbf{making a description of a target class} of objects and in \textbf{detecting whether a new object resembles this class or not}. [\dots] The OCC model is developed using target class samples only.''} \cite{RodionovaOCC:2016}

	\item \e{``In one-class classification framework, an object is classified as belonging or not belonging to a target class, while \textbf{only} sample \textit{examples} of objects \textbf{from the target class} are \textbf{available during the training phase}.''} \cite{JankowskaAVviaCNG:2014}
\end{myitemize}
Note that in the context of \av, the \textbf{target class} refers to the known author $\A$ such that for a document $\Dunk$ of an unknown author $\unknown$ the task is to verify whether $\unknown = \A$ holds.


One of the most important requirements of any existing \AV method is a decision criterion, which aims to accept or reject a questioned authorship. A decision criterion can be expressed through a simple threshold $\Threshold$ or a more complex decision model $\ThresholdModel$. As a consequence of the above statements, the determination of $\Threshold$ or $\ThresholdModel$ has to be performed solely on $\Arefset$, otherwise the \AV method cannot be considered to be unary. However, our conducted literature search regarding existing \AV approaches revealed that there are uncertainties, how to precisely draw the borders between unary and binary \AV methods (for instance, \cite{BoukhaledProbabilisticAV:2014,StamatatosProfileCNG:2014,PothaStamatatosTopicAV:2018}). Nonetheless, few attempts have been made to distinguish both categories from another perspective.
Potha and Stamatatos \cite{PothaStamatatosTopicAV:2018}, for example, categorize \AV methods based on their characteristics being either \textbf{intrinsic} or \textbf{extrinsic} (emphasized by us):
\begin{myenumerate}
	\item \e{``Verification models differ with respect to their view of the task. \textbf{Intrinsic} verification models view it as a \textbf{one-class classification} task [\dots] Such methods [\dots] \textbf{do not require any external resources}.''} \cite{PothaStamatatosTopicAV:2018}

	\item \e{``On the other hand, \textbf{extrinsic} verification models attempt to transform the verification task to a pair classification task by \textbf{considering external documents} to be used as \textbf{samples of the negative class}.}'' \cite{PothaStamatatosTopicAV:2018}
\end{myenumerate}
While we agree with (2), the former statement (1) is unsatisfactory, since intrinsic verification models are \textbf{not necessarily unary}.
The \AV approach GLAD \cite{GLAD:2015}, for instance, directly contradicts the above statement. Here, the authors
\begin{quote}
	\e{``decided to cast the problem as a \textbf{binary classification task} where class values are \classY} [$\A = \unknown$] \e{and \classN} [$\A \neq \unknown$]. [\dots] \e{We do \textbf{not introduce any negative examples} by means of external documents, \textbf{thus} adhering to an \textbf{intrinsic approach}.''} \cite{GLAD:2015}
\end{quote}
A similar contradiction to the statement of Potha and Stamatatos can be observed in the paper of Jankowska et al. \cite{JankowskaCNGAV:2013}, who
introduced the so-called \cng approach that resembles the unary $k$-centers algorithm \cite{TaxOCC:2001}. \cng is intrinsic in that way that it considers only $\Arefset$. On the other hand, the decision criterion which, in this specific case is a threshold $\Threshold$, is determined on a set of verification problems, labeled either as \classY or \classN (``external resources''). Therefore, \cng is in conflict with the unary definition mentioned above. In a subsequent paper, however, the authors refined their \cng approach and introduced an ensemble based on multiple $k$-centers \cite{JankowskaAVviaCNG:2014}. This time, $\Threshold$ was determined solely on the basis of $\Arefset$ such that the modified approach can be considered as a true unary \AV method, according to the aforementioned statements.

In 2004, Koppel and Schler \cite{KoppelAVOneClassClassification:2004} presented the \koppelUnmask approach in their paper \e{``Authorship Verification as a One-Class Classification Problem''}, which, according to the authors, represents a unary \AV method. However, if we take a closer look at the learning process of \koppelUnmask, we can see that it is based on a binary SVM classifier, which consumes feature vectors labeled as \classY and \classN. Here, the task of the SVM is to classify the generated curves according to the two classes \e{same-author} and \e{different-author}. \koppelUnmask, therefore, cannot be considered to be unary, as the decision is not based solely on the documents within $\Arefset$.

It should be highlighted again that these approaches are \textbf{binary and intrinsic} since their decision criteria are determined on a training corpus labeled with \classY and \classN in a binary manner (binary decision regarding problems with known \classY and \classN labels) while regarding the verification they consider, in an intrinsic manner, only $\Arefset$. A crucial aspect, which might have lead to misperceptions regarding the model category of these approaches in the past, is the fact that two different class domains are involved. On the one hand, there is the \textbf{class domain of authors}, where the task is to distinguish $\A$ and $\notA$. On the other hand, there is the \emph{elevated} or \emph{lifted} \textbf{domain of verification problems}, which either falls into class \classY or class \classN. The training phase of binary-intrinsic approaches is used for learning to distinguish these two classes, and the verification task can be understood as putting the verification problem as a whole into class \classY or class \classN, whereby the class domain of authors fades from the spotlight.

In contrast to binary-intrinsic approaches, there exist also \AV approaches that are \textbf{binary and extrinsic} (for example \cite{HernandezHomotopyAV:2015,KhonjiIraqiAV:2014,KoppelWinter2DocsBy1:2014,StamatatosPothaImprovedIM:2017,VeenmanPAN13:2013}) as these methods use external documents during a potentially existing training phase and -- more importantly -- during testing. In these approaches, the decision between $\A$ and $\notA$ is put into the focus, where the external documents aim to construct the counter class $\notA$.

\begin{figure}
	\centering
	\includegraphics[scale=0.58]{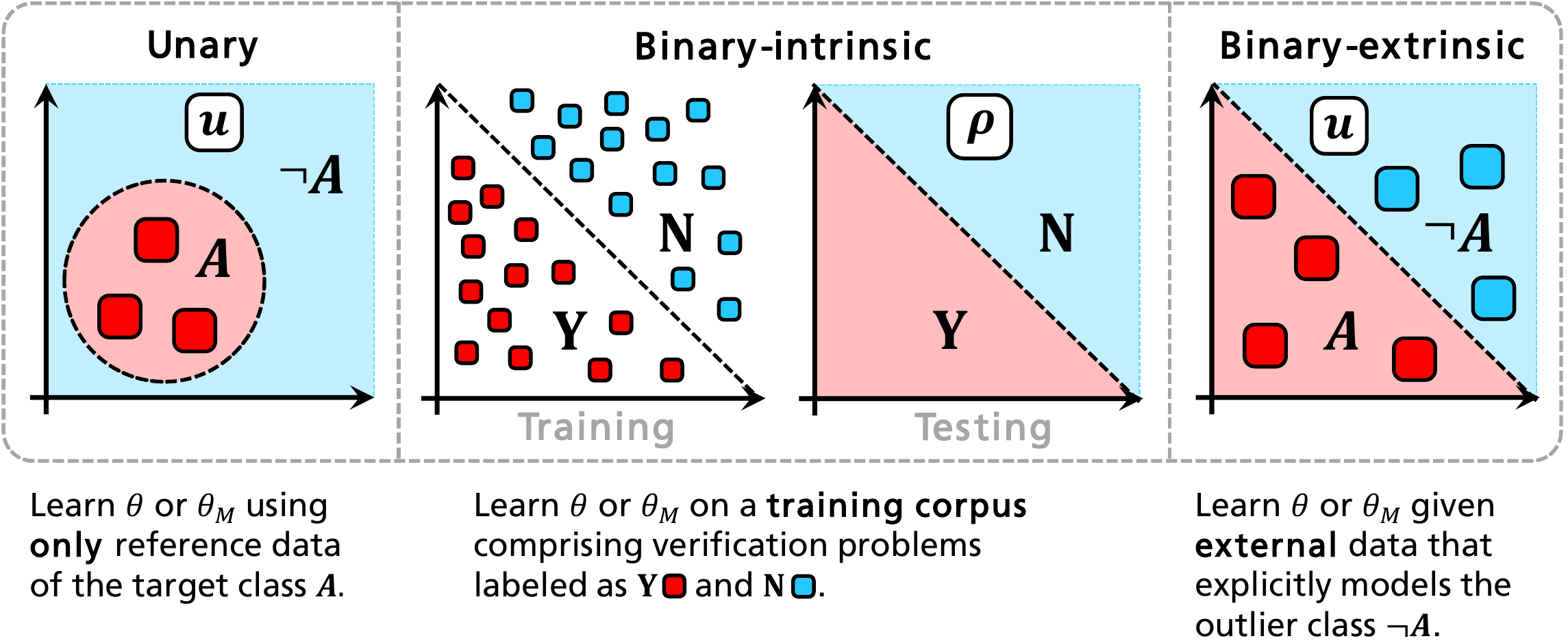}
	\caption{The three possible model categories of \av approaches. Here, $\unknown$ refers to the instance (for example, a document or a feature vector) of the unknown author. $\A$ represents instances of the target class (known author) and $\notA$ the outlier class (any other author). \classY and \classN denote the regions of the feature space where, according to a training corpus, the authorship holds or not. In the  binary-intrinsic case, $\Problem$ denotes the verification problem (subject of classification). \label{fig:AV_Model_Categories}}
\end{figure}

Based on the observations above, we conclude that the \textbf{key requirement} (see illustration in Figure~\ref{fig:AV_Model_Categories}) to judge the model category of an \AV method depends solely on the fact \textbf{how its decision criterion} $\Threshold$ or $\ThresholdModel$ \textbf{is determined}:
\begin{enumerate}
	\item An \AV method is \textbf{unary}, if and only if its decision criterion $\Threshold$ or $\ThresholdModel$ is determined \textbf{solely on the basis of} the target class $\A$. As a consequence, an \AV method cannot be considered to be unary if documents not belonging to $\A$ are used to define $\Threshold$ or $\ThresholdModel$.

	\item An \AV method is \textbf{binary-intrinsic}, if its decision criterion $\Threshold$ or $\ThresholdModel$ is determined on a training corpus comprising verification problems labeled either as \classY or \classN (in other words documents of several authors). However, once the training is completed, a binary-intrinsic method has no access to external documents anymore such that the decision regarding the authorship of $\Dunk$ is made on the basis of the reference data of $\A$ as well as $\Threshold$ or $\ThresholdModel$.

	\item An \AV method is \textbf{binary-extrinsic}, if its decision criterion $\Threshold$ or $\ThresholdModel$ is determined on the basis of external documents that represent the outlier class $\notA$ (the counterpart of $\A$). Here, it is not relevant whether a training corpus was used to optimize $\Threshold$ or $\ThresholdModel$. As long as the method has access to documents of $\notA$, it will remain binary-extrinsic.
\end{enumerate}

It should be highlighted that unary \AV methods (for instance, \cite{HalvaniARES:2014,NealAVviaIsolationForests:2018,StamatatosProfileCNG:2014}) are not excluded to be \textbf{optimizable}. As long as $\Threshold$ or $\ThresholdModel$ is not part of the optimization, the model category of the method remains unary. The rationale behind this is that \textbf{hyperparameters} might influence the resulting performance of a unary \AV method, while the \textbf{decision criterion} itself \textbf{remains unchanged}.

\subsection{Implications}
Each model category has its own implications regarding prerequisites, evaluability, and applicability.

\subsubsection{Unary AV Methods}
One advantage of unary \AV methods is that they do not require a specific document collection strategy to construct the counter class $\neg\A$, which reduces their complexity. Moreover, a training corpus is not required, at least if the method is non-optimizable (for example, \occav \cite{HalvaniOCCAV:2018}. On the downside, the choice of the underlying machine learning model of a unary \AV method is restricted to unary classification algorithms or also unsupervised learning techniques, given a suitable decision criterion. However, a far more \textbf{important implication} of unary \AV approaches concerns their \textbf{performance assessment}. Since unary classification (not necessarily \AV) approaches depend on a fixed decision criterion $\Threshold$ or $\ThresholdModel$, performance measures such as the area under the ROC\footnote{\textbf{R}eceiver \textbf{O}perating \textbf{C}haracteristic.} curve (\auc) are meaningless. Recall that ROC analysis is used for evaluating classifiers, where the decision threshold is not finally fixed. ROC analysis requires that the classifier generates scores, which are comparable across classification problem instances. The ROC curve and the area under this curve is then computed by considering all possible discrimination thresholds for these scores. While unary \AV approaches might produce such scores, \textbf{introducing a variable} $\Threshold$ would \textbf{change the semantics} of these approaches. Since unary \AV approaches have a fixed decision criterion, they provide only a \textbf{single point in the ROC space}. To assess the performance of a unary \AV method it is, therefore, mandatory to consider the confusion matrix that leads to this point in the ROC space.

\subsubsection{Binary AV Methods}
If we design a binary (intrinsic or extrinsic) \AV method, we can choose among a variety of \textbf{binary}\footnote{For example: Support vector machines, logistic regression or perceptron.} and $\bm{n}$\textbf{-ary}\footnote{For example: Naive Bayes, random forests or a variety of neural networks.} classification models. However, if the choice falls on a binary-extrinsic method, a strategy has to be considered, in order to collect representative documents for the outlier class $\notA$. Methods such as \cite{KoppelWinter2DocsBy1:2014,StamatatosPothaImprovedIM:2017,VeenmanPAN13:2013} rely on search engines for retrieving appropriate documents, which might refuse their service if a specified quota is exhausted. Additionally, the retrieved documents make these methods inherently \textbf{non-deterministic}. Moreover, as can be observed in \cite{PANOverviewAV:2013,PANOverviewAV:2014} (as well as in our evaluation in Sec.~\ref{Evaluation}) such methods cause relatively \textbf{high runtimes}. Using search engines also requires an active Internet connection, which might not be available or even allowed in specific scenarios. But even if we can access the Internet to retrieve documents, there is no guarantee that the true author is not among them. With these points in mind, the \textbf{applicability} of binary-extrinsic methods in real-world cases \ie \textbf{forensic settings}, remains questionable. On the other hand, if we consider to design a binary-intrinsic \AV method, it should not be overlooked that the involved classifier learns nothing about individual authors but only similarities or differences that hold in general for \classY and \classN verification problems \cite{KoppelWinter2DocsBy1:2014}.

\section{Evaluation} \label{Evaluation}
Based on our definitions in Sec.~\ref{Analysis}, we investigate the applicability of unary, binary-intrinsic and binary-extrinsic \AV methods. First, we describe which existing \AV methods as well as generic unary classification approaches were considered for our evaluation. Afterwards, we explain which corpora were compiled for the task.

\subsection{Existing AV Approaches}
To assess the performance of \AV methods based on our criteria, we reimplemented 11 existing \AV approaches that have shown their potentials in existing studies as well as in the three PAN \AV competitions from 2013--2015. More precisely, we reimplemented two binary-extrinsic (\koppelGI \cite{SeidmanPAN13:2013} and \veenmanNNCD \cite{VeenmanPAN13:2013}), five binary-intrinsic (\coav \cite{HalvaniARES:2017}, \aveer \cite{HalvaniDFRWS:2016}, \glad \cite{GLAD:2015}, \stamatatosProf \cite{StamatatosProfileCNG:2014} and \koppelUnmask \cite{KoppelAVOneClassClassification:2004}) and four unary \AV approaches (\noeckerDist \cite{NoeckerDistractorlessAV:2012}, \cng \cite{JankowskaAVviaCNG:2014}, \mocc \cite{HalvaniARES:2014} and \occav \cite{HalvaniOCCAV:2018}).

Note that in the original version of both binary-extrinsic approaches \koppelGI and \veenmanNNCD, the authors proposed to use search engine queries to generate impostor documents that are needed to model the counter class $\neg\A$. However, due to quota limits, we decided to use an alternative strategy in our reimplementations. Let $\Corpus = \{ \rho_1, \rho_2, \ldots, \rho_n \}$ denote a corpus. For a given verification problem $\rho_i = ({\Dunk}_i, {\Arefset}_i) \in \Corpus$, we choose all ${\Dunk}_j$ in $\Corpus$ with $i \neq j$ as the impostor set $\mathbb{U}$. However, it should be highlighted that in \koppelGI, the number of impostors is a hyperparameter such that the resulting impostor set is a subset of $\mathbb{U}$, whereas in \veenmanNNCD all $\unknown_j \in \mathbb{U}$ are considered. Although our strategy is not flexible like using a search engine, it has one advantage that here it is assumed\footnote{Note that we cannot be sure if two or more user names in fact refer to the same author.} that the true author of an unknown document is not among the impostors, since in our corpora we know the user names of those who have written all documents.

\subsection{Generic Unary Classification Approaches}
In addition to the reimplemented unary \AV methods, we also considered the four generic unary classification algorithms \occKNN, \occSVM, \occLOF and \occIF (introduced in Sec.~\ref{RelatedWork}) and adapted them to the \AV task. To ensure a fair and equal setting, all classifiers were provided with the same set of features which, according to the literature in \AV and \AA, have been proven to perform very well. The set of features consists of character $n$-grams (with $n\in\{2,3,4\}$), punctuation marks and function words. However, since instead of raw strings the four algorithms require numerical feature vectors as an input, we represent all extracted features according to their relative frequencies in the documents. Instead of selecting the top most frequent features, which is the case in existing \AV approaches such as \cng \cite{JankowskaAVviaCNG:2014} or \stamatatosProf \cite{StamatatosProfileCNG:2014}, we used all occurring features in the texts.
Regarding the two distance-based methods \occKNN and \occLOF, we decided to use the \e{Manhattan} distance, which has been applied successfully in previous authorship analysis studies (for example, \cite{WilliamsAAMalware:2014,EvertUnderstandingDelta4AA:2017,HalvaniDFRWS:2016,JuolaStolermanLingAuthAA:2013,KoppelSeidmanEMNLP:2013}).

\subsection{Corpora}
As a data basis for our evaluation, we compiled two corpora\footnote{All corpora and additional material will be available after publication of this paper. \label{Repository}}. The first corpus represents a collection of 4,000 documents (aggregated postings, crawled from \e{Reddit}) written by 1,000 authors. The second corpus is a collection of 7,000 documents (aggregated product reviews, extracted from the \e{Amazon product data} corpus \cite{AmazonReviewCorpus:2015}) written by 1,400 authors. After aggregating the documents, we split both datasets into training ($\CorpusRedditTrain$, $\CorpusAmazonTrain$) and evaluation corpora ($\CorpusReddit$, $\CorpusAmazon$) and resampled the documents to construct balanced corpora with a bigger number of verification problems. As a result of the resampling procedure, $\CorpusRedditTrain$, $\CorpusReddit$, $\CorpusAmazonTrain$ and $\CorpusAmazon$ ended up with 600, 1,400, 800 and 2,000 verification problems, respectively.

\subsection{Results}
After tuning hyperparameters of all optimizable approaches on the training corpora based on the described training procedure in the respective literature, we applied the learned models together with the non-optimizable methods on both evaluation corpora $\CorpusReddit$ and $\CorpusAmazon$. The results regarding all approaches are listed in Table~\ref{tab:EvaluationResults}. Since we do not limit ourselves to one specific performance measure, we report for each method the outcomes TP, FN, FP, and FN of the corresponding confusion matrix. However, to enable a better comparison, we also list the following ``single number'' evaluation metrics: Accuracy, F$_1$ and Cohen's $\kappa$, where the latter is a relatively new performance measure in the context of \AV, proposed by Halvani et al. in \cite{HalvaniEvalKappa:2018}.

\begin{table}
	\caption{Evaluation results for $\CorpusReddit$ and $\CorpusAmazon$ sorted by $\accuracy$ in descending order. Binary-intrinsic approaches are highlighted by \colorbox{BinaryIntrinisc}{\textbf{purple}}, binary-extrinsic approaches by \colorbox{BinaryExtrinisc}{\textbf{orange\protect\vphantom{l}}} and unary approaches by \colorbox{Unary}{\textbf{green\protect\vphantom{l}}}. \textbf{Non-optimizable} and \textbf{non-deterministic} \AV methods are marked by $\dagger$ and $\star$, respectively.}\label{tab:EvaluationResults}
	\begin{center}
	\begin{tabular}{l@{\hspace{1em}}lrrrrrrrr} \toprule
		& \thead{Method} & \multicolumn{3}{c}{\thead{Performance score}} & \multicolumn{4}{c}{\thead{Confusion Matrix}} & \thead{Runtime}\\
		\cmidrule(r){3-5} \cmidrule(l){6-9} & & \thead{Accuracy} & \thead{$\kappa$} & \thead{F$\bm{_{1}}$} & \thead{TP} & \thead{FN} & \thead{FP} & \thead{TN} & \thead{(hh:mm:ss)} \\\midrule
		\rowcolor{BinaryIntrinisc} \cellcolor{white} & \glad \cite{GLAD:2015} & 0.826 & 0.653 & 0.827 & 579 & 121 &  122 & 578 &  18:06 \\
		\rowcolor{BinaryExtrinisc} \cellcolor{white} & \koppelGI \cite{SeidmanPAN13:2013} ($\star$) & 0.805 & 0.610 & 0.768 &  451 & 249 & 24 & 676 & 5:21:54 \\
		\rowcolor{BinaryIntrinisc} \cellcolor{white} & \aveer \cite{HalvaniDFRWS:2016} & 0.776 & 0.553 & 0.769 &  521 & 179 & 134 & 566 & 0:59 \\
		\rowcolor{BinaryIntrinisc} \cellcolor{white} & \coav \cite{HalvaniARES:2017} & 0.770 & 0.540 & 0.736 &  449 & 251 & 71 & 629 & 0:39 \\
		\rowcolor{Unary} \cellcolor{white} & \occav \cite{HalvaniOCCAV:2018} ($\dagger$) & 0.767 & 0.534 & 0.766 &  533 & 167 & 159 & 541 & 12:27 \\
		\rowcolor{BinaryExtrinisc} \cellcolor{white} & \veenmanNNCD \cite{VeenmanPAN13:2013} ($\dagger$) & 0.764 & 0.529 & 0.695 &  376 & 324 & 6 & 694 & 14:36:25 \\
		\rowcolor{BinaryIntrinisc} \cellcolor{white} & \stamatatosProf \cite{StamatatosProfileCNG:2014} & 0.728 & 0.457 & 0.732 &  519 & 181 & 199 & 501 & 1:55 \\
		\rowcolor{Unary} \cellcolor{white} & \cng \cite{JankowskaAVviaCNG:2014} & 0.719 & 0.437 & 0.743 &  569 & 131 & 263 & 437 & 17:40 \\
		\rowcolor{Unary} \cellcolor{white} & \occLOF \cite{LOFBreunigKriegel:2000} & 0.701 & 0.403 & 0.731 &  568 & 132 & 286 & 414 & 46:11 \\
		\rowcolor{Unary} \cellcolor{white} & \mocc \cite{HalvaniARES:2014} & 0.683 & 0.366 & 0.624 &  368 & 332 & 112 & 588 & 0:57 \\
		\rowcolor{BinaryIntrinisc} \cellcolor{white} & \koppelUnmask \cite{KoppelAVOneClassClassification:2004} ($\star$) & 0.682 & 0.364 & 0.691 &  497 & 203 & 242 & 458 & 7:10 \\
		\rowcolor{Unary} \cellcolor{white} & \occKNN \cite{KhanOCCTaxonomy:2014} & 0.671 & 0.343 & 0.601 &  347 & 353 & 107 & 593 & 52:41 \\
		\rowcolor{Unary} \cellcolor{white} & \occSVM \cite{KhanOCCTaxonomy:2014} & 0.651 & 0.301 & 0.560 &  311 & 389 & 100 & 600 & 1:04:21 \\
		\rowcolor{Unary} \cellcolor{white} & \noeckerDist \cite{NoeckerDistractorlessAV:2012} ($\dagger$) & 0.639 & 0.277 & 0.715 &  634 & 66 & 440 & 260 & \textbf{0:35} \\
		\rowcolor{Unary} \multirow{-15}{*}{\cellcolor{white} \large\rotatebox{90}{$\bm{\CorpusReddit}$}}
		& \occIF \cite{IsolationForestLiu:2008} ($\star$) & 0.501 & 0.001 & 0.612 &  551 & 149 & 550 & 150 & 45:39 \\\midrule
		\rowcolor{BinaryIntrinisc} \cellcolor{white} & \glad \cite{GLAD:2015} & 0.858 & 0.716 & 0.859 &  867 & 133 & 151 & 849 & 16:00 \\
		\rowcolor{BinaryIntrinisc} \cellcolor{white} & \aveer \cite{HalvaniDFRWS:2016} & 0.816 & 0.631 & 0.811 &  790 & 210 & 159 & 841 & 1:39 \\
		\rowcolor{BinaryExtrinisc} \cellcolor{white} & \koppelGI \cite{SeidmanPAN13:2013} ($\star$) & 0.784 & 0.567 & 0.761 &  690 & 310 & 123 & 877 & 52:32 \\
		\rowcolor{BinaryIntrinisc} \cellcolor{white} & \coav \cite{HalvaniARES:2017} & 0.778 & 0.556 & 0.763 &  716 & 284 & 160 & 840 & 2:18 \\
		\rowcolor{Unary} \cellcolor{white} & \occLOF \cite{LOFBreunigKriegel:2000} & 0.769 & 0.537 & 0.779 &  817 & 183 & 280 & 720 & 40:34 \\
		\rowcolor{Unary} \cellcolor{white} & \occav \cite{HalvaniOCCAV:2018} ($\dagger$) & 0.757 & 0.514 & 0.769 &  811 & 189 & 297 & 703 & 12:07 \\
		\rowcolor{Unary} \cellcolor{white} & \occKNN \cite{KhanOCCTaxonomy:2014} & 0.734 & 0.467 & 0.674 &  552 & 448 & 85 & 915 & 41:52 \\
		\rowcolor{BinaryIntrinisc} \cellcolor{white} & \koppelUnmask \cite{KoppelAVOneClassClassification:2004} ($\star$) & 0.731 & 0.462 & 0.728 &  719 & 281 & 257 & 743 & 8:42 \\
		\rowcolor{BinaryIntrinisc} \cellcolor{white} & \stamatatosProf \cite{StamatatosProfileCNG:2014} & 0.722 & 0.443 & 0.719 &  714 & 286 & 271 & 729 & 1:48 \\
		\rowcolor{Unary} \cellcolor{white} & \cng \cite{JankowskaAVviaCNG:2014} & 0.713 & 0.426 & 0.750 &  863 & 137 & 437 & 563 & 23:25 \\
		\rowcolor{Unary} \cellcolor{white} & \mocc \cite{HalvaniARES:2014} & 0.712 & 0.424 & 0.660 &  559 & 441 & 135 & 865 & 1:38 \\
		\rowcolor{Unary} \cellcolor{white} & \occSVM \cite{KhanOCCTaxonomy:2014} & 0.677 & 0.353 & 0.560 &  411 & 589 & 58 & 942 & 1:39:11 \\
		\rowcolor{BinaryExtrinisc} \cellcolor{white} & \veenmanNNCD \cite{VeenmanPAN13:2013} ($\dagger$) & 0.604 & 0.208 & 0.349 &  212 & 788 & 4 & 996 & 15:52:22 \\
		\rowcolor{Unary} \cellcolor{white} & \noeckerDist \cite{NoeckerDistractorlessAV:2012} ($\dagger$) & 0.604 & 0.207 & 0.708 &  960 & 40 & 753 & 247 & \textbf{0:27} \\
		\rowcolor{Unary} \multirow{-15}{*}{\cellcolor{white} \large\rotatebox{90}{$\bm{\CorpusAmazon}$}}
		& \occIF \cite{IsolationForestLiu:2008} ($\star$) & 0.495 & -0.011 & 0.608 &  785 & 215 & 796 & 204 & 36:15 \\\bottomrule
	\end{tabular}
	\end{center}
\end{table}

A variety of observations can be inferred from Table~\ref{tab:EvaluationResults}. In particular, the majority of \textbf{binary-intrinsic} \AV methods tend to \textbf{outperform} both binary-extrinsic and unary approaches. \glad, which is the top performing approach on both corpora, demonstrates that \textbf{binary-intrinsic} approaches are very effective, even though the \AV task itself represents an unary classification problem. The two other \textbf{binary-intrinsic} methods \aveer and \coav also achieve high results, but differ from \glad in several important aspects. \aveer and \coav rely both on simple similarity functions that accept or reject the authorship of unknown documents according to a \textbf{scalar threshold}.\linebreak[3] \glad, on the other hand, is based on an SVM, which is widely known to be a strong classifier. An explanation why \glad is superior might be that the discrimination ability of a single threshold is not fine-granular enough, compared to the hyperplane constructed by the SVM, which separates a 24-dimensional feature space in a non-linear\footnote{\glad utilizes the RBF kernel.} way. Another (or an additional) explanation could be that, in contrast to \aveer and \coav, \glad makes use of several joint features (see Sect.~\ref{ExistingAVApproaches}), which might capture better differences or similarities between the documents.

Furthermore, we can see from Table~\ref{tab:EvaluationResults} that binary-extrinsic approaches also perform very well, in particular \koppelGI. This is consistent with the findings in previous studies such as \cite{JankowskaAVviaCNG:2014,KocherSavoySpatiumL1:2017,StamatatosProfileCNG:2014}. The high results regarding \koppelGI also indicate that considering static corpora to generate impostor documents is a suitable alternative to search engine queries.

When comparing the results of all methods on $\CorpusReddit$ and $\CorpusAmazon$ to each other, we can also see that the majority of the examined \AV approaches perform more or less stable (\glad, \coav, \noeckerDist and \occIF even have exactly the same rankings on both corpora). However, one exception is the binary-extrinsic method \veenmanNNCD, which performs quite well on $\CorpusReddit$, but is among the worst three approaches on the $\CorpusAmazon$ corpus. Unfortunately, there is no clear explanation, if this is caused by the bigger number of available impostors in $\CorpusAmazon$ (here, each ${\Dunk}_i$ is confronted with 400 more impostors than in $\CorpusReddit$) or due to another reason. Therefore, we leave this open question for future work.

From the four examined unary \AV approaches (\noeckerDist, \cng, \mocc and \occav; without the generic unary classification algorithms), \occav yields the best results and performs quite stable on both corpora. Despite of the fact that \occav, which builds on top of \occKNN, belongs to the category of non-optimizable \AV approaches, it seems to generalize very well. This is particularly important in real-world settings such as forensic cases, where training corpora with labeled data of the suspects are not always available.

From all \textbf{generic unary} classification algorithms (\occKNN, \occSVM, \occLOF and \occIF), \occLOF achieves the highest result. One interesting point here is that \occLOF outperforms the closely related \occKNN method, although both not only rely on the same features but also on the same distance function. We wish to highlight at this point that, according to the literature, this is the first time \occLOF has been applied to \AV such that we recommend to investigate its potential in future work.
Another observation that can be seen in Table~\ref{tab:EvaluationResults} is that \occIF performs similar to a random guess. This is noteworthy as recently Neal et al. proposed an \AV method in \cite{NealAVviaIsolationForests:2018} that is very similar to our \occIF implementation\footnote{Our \occIF implementation is mostly based on the \e{scikit-learn} library.}, where the authors report a recognition accuracy exceeding 98\% on the so-called \e{CASIS}\footnote{\textbf{C}enter for \textbf{A}dvanced \textbf{S}tudies in \textbf{I}dentity \textbf{S}cience.} corpus. However, since this corpus is not available online\footnote{An attempt to request the corpus directly from the authors was also not successful.}, we cannot investigate this issue in more detail.

When comparing the unary \AV approaches against the generic unary classification algorithms, there is no clear separation of these to groups regarding their performance since their ranks in Table~\ref{tab:EvaluationResults} are interweaved. There might be a little advantage for the dedicated \AV methods compared to the generic algorithms since \occav is on average the best method within these two groups, and \occIF is clearly separated at the bottom.

Regarding the performance measures listed in Table~\ref{tab:EvaluationResults}, several interesting observations can be made. For example, when looking at the performance results in the \fOne-column, we can see that the ranking of the examined \AV methods differs from those of \accuracy (and $\kappa$). The reason for this can be explained easily, when we consider the underlying formulas of both measures:
\begin{align*}
\textrm{\accuracy} &= \frac{\textrm{TP + TN}}{\textrm{TP + TN + FP + FN}} & \textrm{\fOne} &= \frac{\textrm{2TP}}{\textrm{2TP + FN + FP}}
\end{align*}
The given formula for \fOne is obtained from the given formula for \accuracy by replacing TP + TN with 2TP. Resulting \accuracy values will be \textbf{greater} than \fOne values if TN $>$ TP and \textbf{smaller} if TN $<$ TP holds. This also answers the question, why two \AV methods that perform almost equally in terms of \accuracy (for example, \veenmanNNCD vs. \noeckerDist on the $\CorpusAmazon$) have a significant difference regarding their \fOne values.

The difference between \accuracy and \fOne is more than a matter of interchangeable design choices. The design of the \fOne measure leads to the problem that resulting \fOne values can be quite misleading. For instance, if an \AV method predicts always \classY (\eg due to a weak threshold), \fOne will result in \nicefrac{2}{3} on a balanced corpus. In contrast, \accuracy will result in \nicefrac{1}{2}, which can be interpreted as a \textbf{coin toss}. In the case of an \AV method that predicts always \classN, \fOne will be 0 while \accuracy will result in \nicefrac{1}{2} again.

Putting the discussion to a more abstract level, the problem is that the measure \fOne ignores the true negatives (TN) in contrast to \accuracy (and $\kappa$). Ignoring TN is generally not reasonable in the context of \AV, as it must be measurable if a method is able to correctly predict such cases, where the authorship does not hold. Based on these findings, we discourage to use \fOne for assessing the performance of \AV methods.

An observation regarding \accuracy and $\kappa$ can be made when comparing the columns for \accuracy and $\kappa$ in Table~\ref{tab:EvaluationResults} to each other. Both measures preserve the same ranking, and a closer look reveals that $\kappa$ has a linear relationship to \accuracy on balanced corpora (such as $\CorpusReddit$ and $\CorpusAmazon$). The explanation for this can be shown based on the definition of $\kappa$:
\begin{align*}
n      &= \textrm{TP + FN + FP + TN} \\
p_0    &= n^{-1}(\textrm{TP+TN}) \\
p_c    &= n^{-2} \left( (\textrm{TP+FN})(\textrm{TP+FP})+(\textrm{FP+TN})(\textrm{FN+TN}) \right) \\
\kappa &= \frac{p_0 - p_c}{1 - p_c}
\end{align*}
For \textbf{balanced} corpora, $p_c$ results in 0.5 such that $\kappa = 2 \times \textrm{\accuracy} - 1$ holds. However, in cases where corpora are imbalanced, it makes more sense to use $\kappa$ instead of \accuracy, as the latter favors the majority class. A visual inspection of the behavior of both measures regarding imbalanced corpora is given in \cite{HalvaniEvalKappa:2018}.

A closer look on the last column in Table~\ref{tab:EvaluationResults} also reveals a number of issues that may require some consideration. Compared to the binary-intrinsic \AV methods, the majority of the unary approaches obviously require more runtime. One exception here is \noeckerDist, which needs on average $\approx$ 31 seconds to process a whole test corpus. Binary-extrinsic approaches require even more runtime, compared to almost all unary approaches. A good trade-off between performance and runtime (which might be an important issue in the context of an IR system) can be observed for \aveer, followed by \coav.

\section{Conclusion and Future Work} \label{Conclusions}
Based on a comprehensive literature review of numerous \AV studies, we identified a number of misunderstandings regarding the different model categories of existing \AV approaches, which have serious implications regarding their prerequisites, evaluability, and applicability. We defined clear criteria that aim to draw precise borders between the different categories of \AV approaches and explained, which challenges occur in terms of evaluation, when an \AV method is based on a \textbf{fixed} decision criterion. 
Given our definitions, we reimplemented a number of existing unary, binary-intrinsic and binary-extrinsic \AV methods and assessed their performance on two large self-compiled corpora, which we made available for the \AV community. One of our observations was that specific unary \AV methods can not only outperform their binary-intrinsic and binary-extrinsic counterparts but also perform stable across the different corpora. 
We have shown why the \fOne performance measure can be misleading in the context of \av and also highlighted the connection between \accuracy and $\kappa$, which occurs when the considered corpora are balanced.  

Furthermore, we tested the applicability of four generic unary classification algorithms for the \AV task, where all four were given exactly the same feature vectors (and in two cases the same distance function). It turned out that distance-based unary classifiers are able to outperform existing \AV methods and achieve (at least partially) promising results. For the first time, we applied \occLOF in the context of \AV, which not only outperformed \occSVM (a commonly used baseline in existing \AV studies) but also requires less runtime. Therefore, we recommend to consider \occLOF as a starting point for future \AV approaches.

In the near future, we will expand our evaluation on more corpora and organize the field of \av in more depth, through the definition of additional properties such as \textbf{reliability}, \textbf{robustness} and \textbf{interpretability}. Here, especially the latter is gaining more and more importance. Moreover, we plan to compile additional corpora in order to investigate the question, if the findings in this paper also hold in other corpora, which differ in terms of topic, genre and the language itself.
\bibliographystyle{plain}
\bibliography{Bibliography}
\end{document}